\documentclass[journal=aamick,manuscript=article,reqno]{amsart}
\usepackage{empheq}

\usepackage[version=3]{mhchem} 
\usepackage[T1]{fontenc}  
\usepackage[super]{natbib}
\usepackage{physics}
\calclayout
\usepackage{cancel}
\usepackage[foot]{amsaddr}
\usepackage{subfig}
\usepackage{mathtools}
\usepackage{amssymb} 
\usepackage{gensymb}
\usepackage{graphicx}
\usepackage{filecontents}
\usepackage{bm}
\usepackage{xr}
\usepackage{booktabs}
\usepackage{siunitx}
\usepackage[svgnames, table]{xcolor}
\usepackage{array, cellspace, makecell}
\usepackage{float}

\usepackage{enumitem}
\newlist{tabitem}{itemize}{1}
\setlist[tabitem]{wide=0pt, nosep, leftmargin= * ,label=\textbullet,after=\vspace{-\baselineskip},before=\vspace{-0.6\baselineskip}}

\newcommand{\beginsupplement}{%
        \setcounter{section}{0}
        \renewcommand{\thesection}{S\arabic{section}}%
        \setcounter{table}{0}
        \renewcommand{\thetable}{S\arabic{table}}%
        \setcounter{figure}{0}
        \renewcommand{\thefigure}{S\arabic{figure}}%
        \setcounter{equation}{0}
        \renewcommand{\theequation}{S\arabic{equation}}
     }
\usepackage{abstract}

\usepackage{tabularx}

\usepackage{ltablex}
\usepackage{arydshln}
\keepXColumns
\newcolumntype{L}{>{\centering}X}

\title{The physical origin of aneurysm growth, dissection, and rupture}
  
\author{Tom Y. Zhao$^{\dagger,\ominus,\S}$ | PhD}
\author{Jin-Tae Kim$^{\oplus,\ominus,\uplus}$ | PhD}
\author{Min Cho$^{\otimes}$}
\author{Akhil Narang$^{\ddagger}$ | MD}
\author{John A. Rogers$^{\oplus}$ | PhD}
\author{Neelesh A. Patankar$^{\dagger,\S}$ | PhD}

\address[$\dagger$]{Northwestern University, Department of Mechanical Engineering: 2145 Sheridan Road, Evanston, Illinois 60208, USA}
\address[$\oplus$]{Northwestern University, Querrey Simpson Institute for Bioelectronics: 11th Floor, 303 E. Superior Chicago, IL 60611, USA}
\address[$\uplus$]{Department of Mechanical Engineering, Pohang University of Science and Technology, Pohang 37673, Republic of Korea }
\address[$\otimes$]{University of Illinois Urbana-Champaign, Department of Chemical and Biomolecular Engineerings: Urbana, IL 61801, USA}
\address[$\ddagger$]{Northwestern University Feinberg School of Medicine, 420E Superior St, Chicago, IL 60611, USA}
\address[$\ominus$]{These authors contributed equally to this work}

\address[$\S$]{Corresponding author: n-patankar@northwestern.edu}

\begin{document}

\maketitle

\begin{abstract}
Rupture of aortic aneurysms is by far the most fatal heart disease, with a mortality rate exceeding 80\%. There are no reliable clinical protocols to predict growth, dissection, and rupture because the fundamental physics driving aneurysm progression is unknown. Here, via in-vitro experiments, we show that a blood-wall, fluttering instability manifests in synthetic arteries under pulsatile forcing. We establish a phase space to prove that the transition from stable flow to unstable aortic flutter is accurately predicted by a flutter instability parameter derived from first principles. Time resolved strain maps of the evolving system reveal the dynamical characteristics of aortic flutter that drive aneurysm progression. We show that low level instability can trigger permanent aortic growth, even in the absence of material remodeling. Sufficiently large flutter beyond a secondary threshold localizes strain in the walls to the length scale clinically observed in aortic dissection. Lastly, significant physical flutter beyond a tertiary threshold can ultimately induce aneurysm rupture via failure modes reported from necropsy. Resolving the fundamental physics of aneurysm progression directly leads to clinical protocols that forecast growth as well as intercept dissection and rupture by pinpointing their physical origin.

\end{abstract}

\section{Introduction}
The global number of deaths caused by aortic aneurysms increased by $29\%$ over the ten year period from 2009\cite{2022Wang, 2013Li, 2016Mathur}. However, the fundamental cause of aneurysm progression remains unknown.

This hampers both monitoring and management of aneurysms over time \cite{2014Geisbusch}. Without an understanding of the root mechanism driving aneurysm development, the clinical standard remains purely retrospective, in that growth is tracked by taking the difference in aortic diameters measured one or more years apart \cite{2015McLarty}. Clinical decision making can thus only occur in hindsight, often after abnormal growth, dissection, or rupture have already occurred \cite{2018Stout}. Since there are no effective prevention or treatment strategies \cite{2020Wang}, surgery remains the only clinically proven method of intervention. This places an additional burden on accurately forecasting aneurysm outcomes to weigh the risk of surgical repair against disease progression.

The state of the art in predicting aneurysm growth relies on regression analysis and machine learning \cite{2020Chandrashekar, 2020Hirata}. Both approaches are constrained by the size, diversity, and availability of the data sampled. Other than the current disease state, many potentially confounding factors such as age, sex, smoking history, congenital diseases, aortic regurgitation, ejection fraction < 50\%, etc. are typically incorporated as input \cite{2017Cabitza}; without an understanding of the core mechanism underlying aneurysm development, every conceivable component, from physiological feature to demographic information, must be included to wring out predictive capability.

To address this gap, we propose that the physics of aneurysm progression originates from a blood-aortic wall instability that occurs under pulsatile forcing of the human heart \cite{2021Zhao}. The transition from stable flow to parametric instability can be pinpointed via a linear stability analysis of blood flow through the aorta. This ab-initio derivation yields the so-called flutter instability parameter, a dimensionless number minus its critical threshold that signals the onset of the "fluttering' instability in a blood vessel \cite{2021Zhao}. 

The flutter instability parameter is a function of physiological properties like pulse wave velocity, pressure driven flow acceleration, blood viscosity, aortic area, pulse rate, etc. The aortic flutter it describes is akin to the "flutter' mode of a banner in the wind, where the physical stiffness, wind acceleration, air viscosity, banner area, wind pulsation, etc. play analogous roles to their physiological counterparts. No machine learning or statistical correlations are needed as input since the flutter instability parameter is a purely analytical solution.

From the proposed theory, we hypothesize that the flutter instability parameter (\textbf{1}) accurately captures the transition from stable flow to unstable fluttering. We also conjecture that the resulting aortic flutter (\textbf{2a}) either triggers the onset of abnormal growth by inducing significant local stresses and strains in the aortic wall, or (\textbf{2b}) signals growth because flutter accompanies some "ground truth" mechanism responsible for abnormal dilatation. In prior work\cite{2021Zhao}, we showed that the aneurysm physiomarker measured from 4D flow magnetic resonance imaging (4D flow MRI) accurately forecast aneurysmal dilatation for each patient at follow-up. Although the clinical utility of this flutter instability parameter as an aneurysm physiomarker is demonstrated, the underlying assumptions (\textbf{1}) and (\textbf{2}) have not been physically validated outside of theory.

\begin{figure}[]
\centering
\includegraphics[width=10cm]{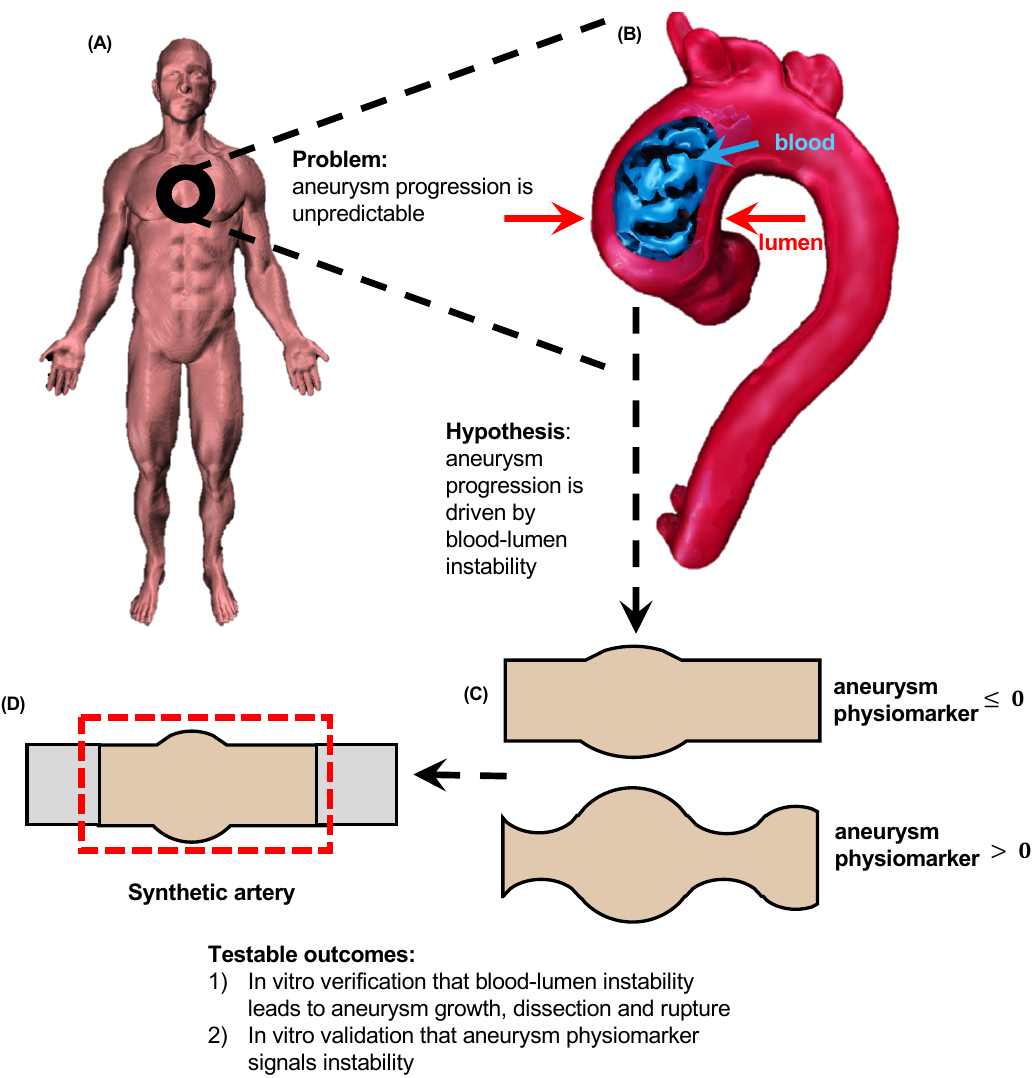}
\caption{Aneurysm physiomarker forecasts aneurysm growth, dissection and rupture by signaling transition from stable flow to fluid-structure instability. Schematic illustrations of (A) human aneurysm growth in (B) the thoracic aorta. Progression of aortic aneurysms is hypothesized to occur via a fluid-structure instability predicted by ab-initio analysis of the underlying biophysics. This analysis also yields an aneurysm physiomarker (C) which signals the onset of the proposed fluid-structure instability. Two testable outcomes can be resolved in the present work by using an in-vitro experimental system that drives pulsatile flow through synthetic arteries. It is shown that (D) the aneurysm physiomarker signals instability onset, and that the instability can lead to permanent arterial growth, dissection and ultimately rupture.} 
\label{fig: problem_setup}
\end{figure}

Here, we employ in vitro experiments to prove that the underlying physical conjectures comprising the aneurysm physiomarker are correct. Figure \ref{fig: problem_setup} sets up this problem; we validate that the aneurysm physiomarker accurately signals the onset of the fluid-structure instability (\textbf{1}) as well as verify that this instability can directly lead to aneurysm growth, dissection and rupture (\textbf{2a}). 

Two clinical outcomes stem from this work. First, we show that flutter can be distinctly visualized via imaging. This suggests echocardiagrams may be used to measure flutter in-vivo, which offers an alternate modality to track aneurysm progression. Secondly, we observe that when the aneurysm physiomarker exceeds a secondary, empirically measured threshold, strain patterns localize along the synthetic artery (SA) in the length scale associated with clinical aortic dissection. A tertiary, empirically measured threshold can also be identified that marks the onset of rupture for the synthetic artery. 

As the physical underpinnings of the flutter instability parameter have been validated, it can be used as a measurable, treatable, and predictive surrogate for aneurysm progression. Its component physiological properties can also be managed to optimally lower the aneurysm physiomarker value toward stable flow for each patient, thereby forestalling dissection and rupture indefinitely.

\section{Overview of experimental equipment, procedures, imaging techniques} 

The experimental setup includes a series of controlled synthetic arteries with various elastic properties \cite{ma2018} and custom artificial heart system that consists of two cylinders as right atrium (RA) and right ventricle (RV), two prosthetic heart valves as pulmonary (PV) and tricuspid (TV) valves, as well as commercial flow (TS410 tubing module, Transonic Systems Inc.) and pressure sensors (Tru-Wave Disposable Pressure Transducer, Edwards Lifesciences) \cite{kwon2023}. Three cases of SAs are prepared with 10:1, 20:1, 30:1 (the base to curing agent) mixing ratios of Polydimethylsiloxane (PDMS) using 3D printed molds for the dimension of a tube, similar to an actual aorta's (diameter $D=2.5$ cm, thickness $t=1.8$ mm and length $L=12$ cm). The PDMS poured into the mold are cured overnight in the 70$^{\circ}$C oven. The resulting moduli for 10:1, 20:1 and 30:1 SA measured by a force gauge sensor (ME-100, Mark-10) with the motorized tension test stand (ESM303, Mark-10) are 2.51, 0.91 and 0.18 MPa (Fig.2C), consistent with the literature values \cite{wang2014}.

A mechanical pump inside the RA cylinder generates flows through the SAs with controllable frequency and amplitude of the flow rate and pressure monitored by the integrated flow and pressure sensors. In-vitro characterization of aortic instability is investigated by using a combination of optical approaches including Particle Tracking Velocimetry (PTV) to track fiducial marks on the surface of artificial arteries and three dimensional Digital Image Correlation (3D-DIC) to quantify the 3D surface deformations of SAs \cite{kim2022,solav2018}.  For PIV experiments, test cases of 3 BPMs (60, 105 and 150) with 8 different amplitudes of the oscillating pump for 10:1 and 20:1 AA as well as 6 different amplitudes for 30:1 AA, total 66 cases, are collected with a 2MP camera operating at the sampling rate of 240Hz. Out of 66 cases, 10 representative cases are chosen for additional DIC experiments with two synchronized 2MP cameras (2048 $\times$ 1088; HT-2000M, Emergent) with 35-mm imaging lenses (F1.4 manual focus; Kowa) operating at the sample rate of 500 fps. Extrinsic and intrinsic camera parameters are optimized to minimize errors associated with the 3D reconstruction process. The investigation volume was $100 \times 80 \times 60$ mm$^3$ with the RMSE value of 0.45 mm between true calibration and 3D reconstructed points (Fig. 2B). SAs are uniformly coated with black dots ($\sim$ 200-500 $\mu$m) using a spray painting technique \cite{ni2022}. The DIC subset radius and spacing are set as 20 and 10 pixels over 3400 points. The Triangular Cosserat Point Elements (TCPE) method is implemented for quantifying strain distributions associated with the aortic instability \cite{kim2023mechanics}. The method is particularly useful for separating the rigid body motion and computing the finite nonlinear strain field by treating each tetrahedron as a Cosserat Point Element \cite{solav2014}

\subsection{Temporal harmonic ratio}
The temporal harmonic ratios are calculated as the power of the fundamental frequency (equivalent to the pulsatile frequency set by the pump) divided by the summed power of all subsequent, larger frequencies (inclusive of the fundamental frequency). Small values of this temporal harmonic ratio correspond to a response dominated by higher order harmonic and subharmonic modes to the driving frequency; by definition, this signals the presence of a Floquet-type instability. This analysis was performed for both DIC and PTV experiments due to sufficient temporal resolution in both modalities. 

Of the three SA cases, the 10:1 mixing ratio produced walls that were too stiff; the signal to noise ratio in the frequency space was too small for usable data to be extracted under pulsatile forcing. Specifically, any signal where the power at the fundamental frequency < 1 mm$^2$ was excluded from analysis. The two lowest gain loading scenarios for the 20:1 mixing ratio were likewise dropped based on this criterion.
 
\subsection{Spatial harmonic ratio}
The spatial harmonic ratios were calculated as the power of the DC frequency divided by the summed power of all subsequent, larger modes (inclusive of the DC component). The DC component was chosen because the system was not driven at a specific spatial wavelength by the operating pump. Small values of this spatial harmonic ratio correspond to a response dominated by higher order modes, which implies concentration of local strain patterns in smaller wavelengths rather than a uniform response throughout the synthetic artery. This analysis was performed for only the DIC experiments, since sufficient spatial resolution is only available through DIC. 
\subsection{Aneurysm physiomarker}
The aneurysm physiomarker was derived \cite{2021Zhao} in a linear stability analysis of the 1D conservation laws for momentum and mass for pulsatile flow through a tubular vessel. The system was closed by a linear tube law, but generalizes to any constitutive model describing the relationship between pressure gradient and lumen area via the use of the pulse wave velocity \cite{2021Zhao}. This ab-initio derivation yields the following dimensionless number encapsulating the important drivers and inhibitors of the flutter instability
\begin{equation}
N_{\omega} = \frac{\bar{\phi}A_m^{1/2}}{(\frac{\beta_b}{2}\pi\nu)c_{pw}}.
\end{equation}
Here, $\bar{\phi}$ is the pressure gradient normalized by blood density, $A_m$ is the local aortic cross-sectional area, $\beta_b$ is the wall shear coefficient dependent on the pulsatility of the flow\cite{2021Zhao}, $\nu$ is the blood kinematic viscosity, and $c_{pw}$ is pulse wave velocity of the blood vessel. The difference between this dimensionless number and its critical threshold at marginal stability is the flutter instability parameter that describes instability onset
\begin{equation}
N_{\omega,\text{sp}} = N_{\omega}-N_{\omega,\text{threshold}},
\end{equation}
where $N_{\omega,\text{threshold}}$ is the global minimum along the marginal stability curve separating the zone of stability (perturbations damp out) from the instability regime (perturbations glow). It can be found by solving a matrix equation, as explicitly detailed in Zhao et al. \cite{2021Zhao}. Thus for $N_{\omega,\text{sp}}>0$, the pulsatile instability is triggered by the unbounded growth of perturbations; for $N_{\omega,\text{sp}}<0$, flow is stable since perturbations decay. We have observed in prior work \cite{2021Zhao} that the flutter instability parameter can be used as an aneurysm physiomarker to accurately forecast abnormal aneurysm growth. 
\section{Results and Discussion}
\begin{figure}[]
\centering
\includegraphics[width=10cm]{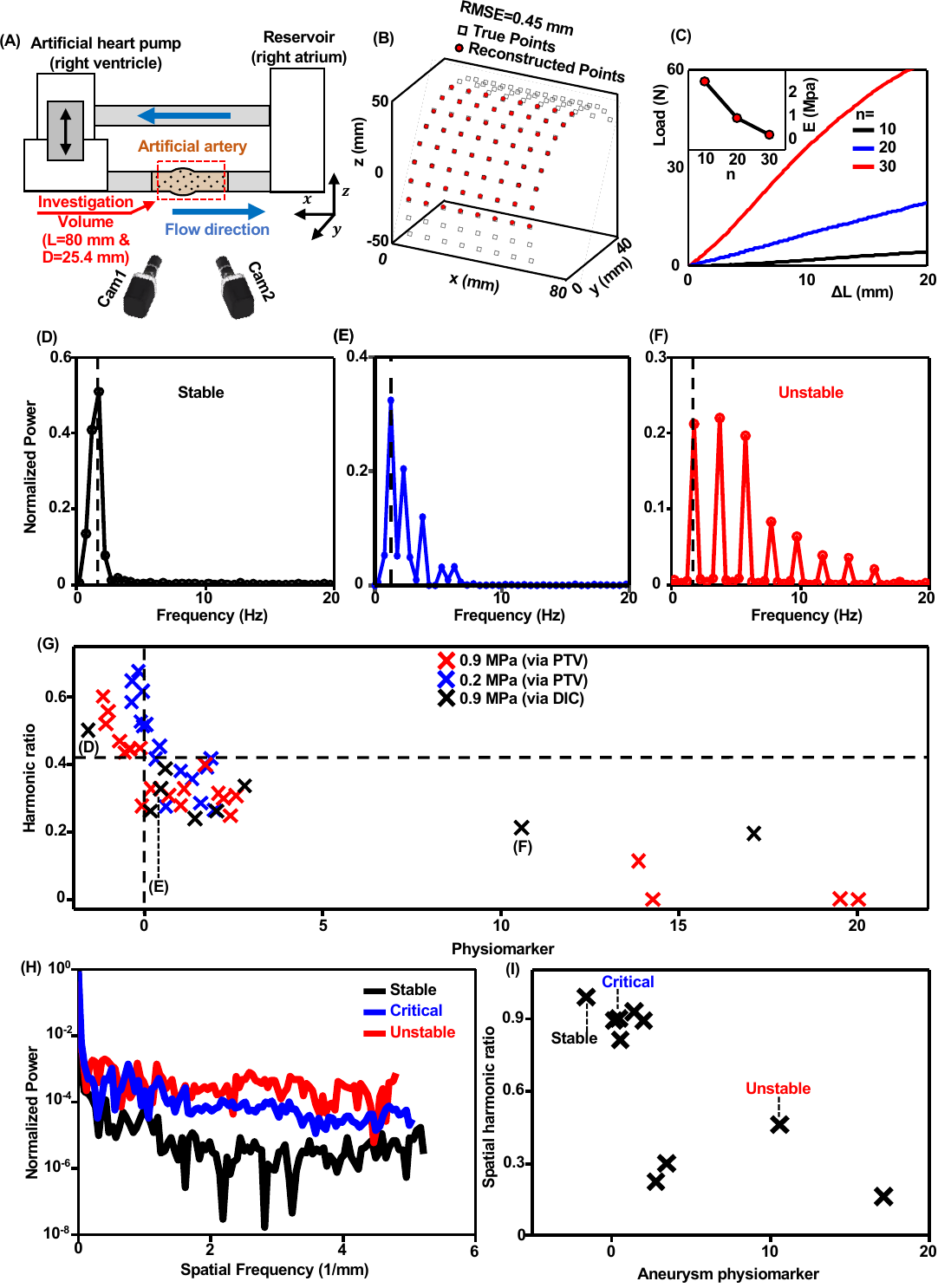}
\caption{Overview of experimental results. The (A) schematic illustration of the experimental setup and (B) 3D DIC validation: true calibration points (black squares) vs reconstructed points (red crosses). RMSE between 66 reconstructed and actual calibration points is 0.45 mm. (C) Load vs deformation curves for the synthetic arteries under tension tests were performed to extract the elastic moduli for three  base to curing agent ratios $n$. The Fourier spectrum in time of a stable case with aneurysm physiomarker $N_{\omega,\text{sp}}=-1.587<0$ (D), critical case $N_{\omega,\text{sp}}=0.448\approx0$ (E), and unstable case $N_{\omega,\text{sp}}=10.569>0$ (F) were visualized as a sanity check. From these spectra, the temporal harmonic ratio (G) were extracted as the power of the fundamental frequency divided by the summed power of all higher order modes (inclusive of the fundamental frequency) and plotted against the measured aneurysm physiomarker. The Fourier spectrum in space (H) were also calculated to report the spatial harmonic ratio (I) on a log scale, which were calculated as the power of the DC component divided by the summed power of all higher order modes (inclusive of DC component).} 
\label{fig: exp_results}
\end{figure}
\subsection{In-vitro validation of the fluid-structure instability}
By sweeping the parameter space of driving frequencies $\omega$ and pressure differentials $\Delta p$ of the pump, we establish the phase diagram of the temporal harmonic ratio vs the aneurysm physiomarker $N_{\omega,\text{sp}}$ shown in Figure \ref{fig: exp_results}\textbf{G}. A clear trend emerges; the temporal harmonic ratio (where smaller values signal flutter) varies inversely with the aneurysm physiomarker (where larger values $>0$ signal flutter). 

Moreover, the analytical threshold built into the aneurysm physiomarker by definition \cite{2021Zhao} sets the critical point to 0. That is, the inception of flutter should occur theoretically at aneurysm physiomarker $N_{\omega,\text{sp}}= 0$. We note from Fig. \ref{fig: exp_results}\textbf{G} that the temporal harmonic ratio drops sharply at this critical point. If an empirically observed optimal value of temporal harmonic ratio $=0.42$ is chosen to divide the phase space into four sectors along with $N_{\omega,\text{sp}}= 0$, the effective confusion matrix yields a sensitivity of 0.97, a specificity of 0.90, and an accuracy of 0.94. The area under the curve (AUC) of a receiver operating characteristic (ROC) analysis is 0.99. These values demonstrate a distinct jump in the temporal harmonic ratio at $N_{\omega,\text{sp}}=0$.

From the phase diagram, we extract three representative cases  with aneurysm physiomarker values $-1.587$ ($<0$, stable, Fig. \ref{fig: exp_results}\textbf{D}), $0.448$ ($\approx 0$, critical, Fig. \ref{fig: exp_results}\textbf{E}), and $10.569$ ($>0$, unstable, Fig. \ref{fig: exp_results}\textbf{F}). As expected, the stable case is dominated by the fundamental response with minimal higher frequency noise. Meanwhile, the critical case exhibits a distinct subharmonic peak to the fundamental driving frequency, which nonetheless provides the bulk contribution to the power spectrum. Lastly, the unstable case yields the reverse scenario where most of the spectral power is partitioned to higher order harmonic modes. 

These scenarios, in addition to DIC and PTV cases presented in Fig. \ref{fig: DIC_panel} and Fig \ref{fig: PTV_panel}, provide a sanity test demonstrating that the jump in temporal harmonic ratios indeed reflects a transition to aortic instability at $N_{\omega,\text{sp}}=0$. Thus, the theoretically predicted fluid-structure instability has been observed in experiment, and the aneurysm physiomarker derived to measure the binary presence of the instability has likewise been physically validated. 

\subsection{Verification that instability leads to aneurysm growth, rupture, or dissection}
The important question is whether this observed fluid-structure instability actually causes permanent arterial dilatation, rupture or dissection. We also seek to elucidate the mechanism by which these modes occur. 
\subsubsection{Dilatation}
Figure \ref{fig: permdef} shows that the maximum radial deformation measured via DIC while fluid is pumped through the synthetic arteries trends proportionally with increasing value of the aneurysm physiomarker. This is unsurprising as greater $N_{\omega,\text{sp}}>0$ indicates increasing aortic instability as power shifts away from the fundamental frequency toward higher harmonic modes (Fig. \ref{fig: exp_results}\textbf{G}). 

The maximal radial deformation comprises both the elastic and plastic response of the hyperelastic material. We also separate out the plastic radial deformation in Fig. \ref{fig: permdef} by comparing the relaxed configuration of the synthetic artery after each experiment (no further pumping, and excess fluid allowed to evacuate test section) to its initial condition. Note that permanent deformation does not trend near monotonically with greater aneurysm physiomarker value, but instead has a nonnegligible dependence on the strain history of the tube; each labeled number corresponds to the order in which these experiments were performed. 

Specifically, trial one (Fig. \ref{fig: exp_results}\textbf{D}) resulted in $N_{\omega,\text{sp}}\approx 0$ near the critical threshold of instability onset. This yielded no permanent deformation despite moderate maximum deformation during pumping. Yet trial five (Fig. \ref{fig: exp_results}\textbf{A}) was a stable case with large harmonic ratio, low max deformation, and $N_{\omega,\text{sp}}<0$ that nonetheless produced nonzero permanent deformation. This suggests that once the instability threshold has been triggered, even a return to stable ($N_{\omega,\text{sp}}< 0$, trial five) or minimally unstable ($N_{\omega,\text{sp}}\approx 0$, trial three, six) flow is sufficient to induce steady growth. Our observation agrees with the low, but present growth rate even for small aneurysms (0.08 cm/year) that are not at great risk of dissection or rupture. These experiments show that sustained aneurysm growth can result via the observed fluid-structure instability after the threshold $N_{\omega,\text{sp}}=0$ is breached in the loading history of the artery, even in the absence of tissue remodelling.

\subsubsection{Dissection}

A classic example of aortic dissection begins as a localized tear in the inner aortic wall (intima) \cite{2007Chirillo}. This permits blood to flow through the dissected intimal flap into the layers between the intima and media, forming a false lumen. Without surgical intervention or ameliorating factors such as downstream communication between the false and true lumen to permit outflow, the true lumen may ultimately collapse \cite{2003Quint}. 

Fig. \ref{fig: exp_results}\textbf{H} and \ref{fig: exp_results}\textbf{I} show that the spatial harmonic ratio measured via DIC scales inversely with the aneurysm physiomarker. That is, increasing instability not only results in higher temporal harmonics, but also induces a shift in power from an average DC spatial response toward higher spatial frequency modes. This corresponds to a sharp localization in strain associated with a steep drop in the spatial harmonic ratio near $N_{\omega,\text{sp}}\approx 2$ and is confirmed by the strain concentration observed in Fig. \ref{fig: results2}\textbf{F}.  

Quantitatively, we can estimate the length scale of strain localization by picking the first peak in the Fourier spectrum after the DC component. The first peak occurs at 1.34 mm, which is in the same order as clinically observed, minimum length scale of 1 mm for intimal tears \cite{2003Quint, 2020Li}. This flutter induced strain concentration at sufficiently large values of $N_{\omega,\text{sp}}$ may directly weaken the intima to create tears, or cause local aortic wall hematoma to drive secondary intimal tearing. 

\subsubsection{Rupture}

We observed two rupture cases corresponding to $N^{\text{rupture, 1}}_{\omega,\text{sp}}=12.47$ (DIC, 20:1 ratio) and $N^{\text{rupture, 2}}_{\omega,\text{sp}}=15.33$ (PTV, 30:1 ratio). Both scenarios arose for a pulse rate of 105 bpm. The first failure case with $N^{\text{rupture, 1}}_{\omega,\text{sp}}=12.47$ exhibited a pinhole rupture (Fig. \ref{fig: rupture_results}\textbf{D}), which is a clinical failure mode that has been observed for aortic root aneurysms \cite{2004Sako}. In the case study, both patients had elevated pulse rates in the same range as our in-vitro system (110 and 120 bpm).

The second mode of failure involved a longitudinal slit "unzipping" along the length of synthetic artery (Fig. \ref{fig: rupture_results}\textbf{J}). We note that these "unzipped" slits have likewise been observed \cite{2019Okada, 1955Parkhurst} through necropsy and occured on the order of $5$ mm along the longitudinal direction of blood flow.  

Fig. \ref{fig: rupture_results}\textbf{I} shows that rupture appears to follow a run-away reaction in which radial tube deformation increases linearly over each cycle. That is, fluid accumulates along the test section of the synthetic artery since the impedance associated with the distal pressure is much larger than that corresponding to the expanding plastic stress of the wall. That both final failure modes observed in vitro match rupture characteristics found in vivo (especially the pinhole failure) suggest that this irreversible buildup of fluid and pressure along the dilated cross-section occurs in the last moments of the aneurysm lifecycle at large, positive magnitudes of the aneurysm physiomarker. 

\begin{figure}[]
\centering
\includegraphics[width=10cm]{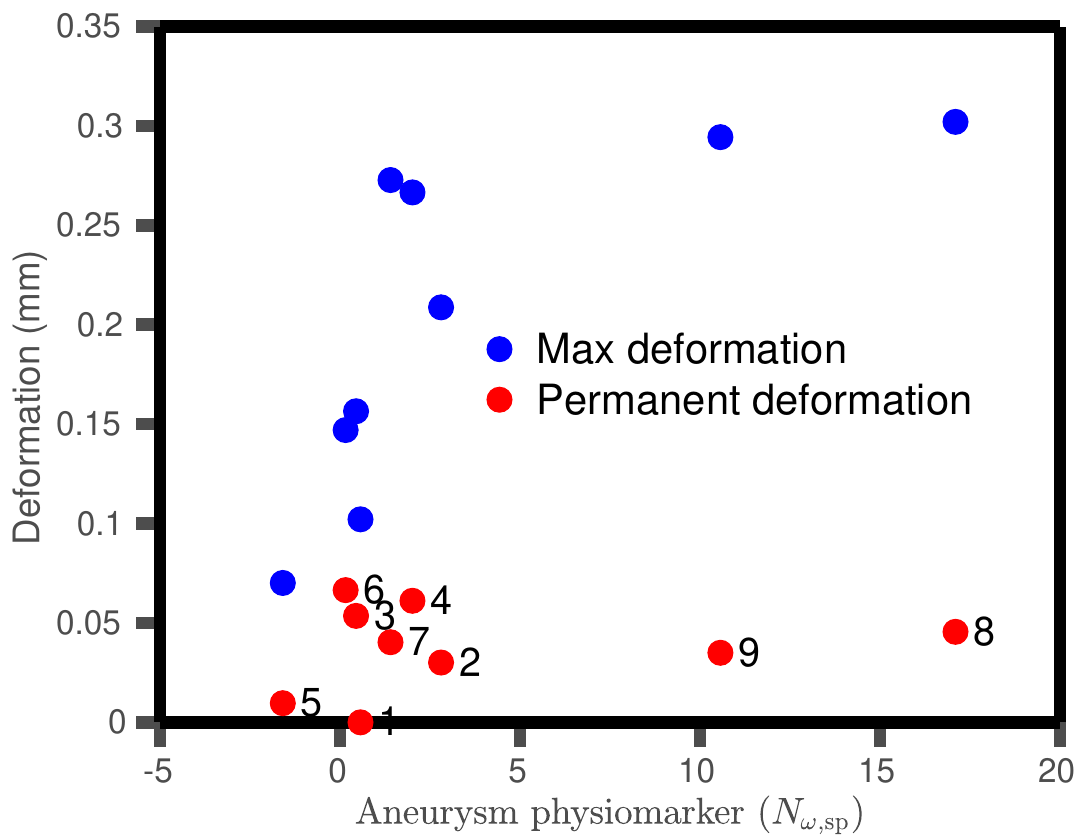}
\caption{The max (elastic+plastic) and permanent plastic deformation for each DIC case. For each scenario corresponding to a different value of the aneurysm physiomarker $N_{\omega,\text{sp}}$, blue data points give the maximum radial deformation over multiple pumping cycles with respect to the initial configuration of that case. Red data points report the maximum plastic deformation measured after pumping has ceased and excess fluid is allowed to evacuate the test section, relative to the initial configuration of that case.}
\label{fig: permdef}
\end{figure}

\begin{figure}[]
\centering
\includegraphics[width=10cm]{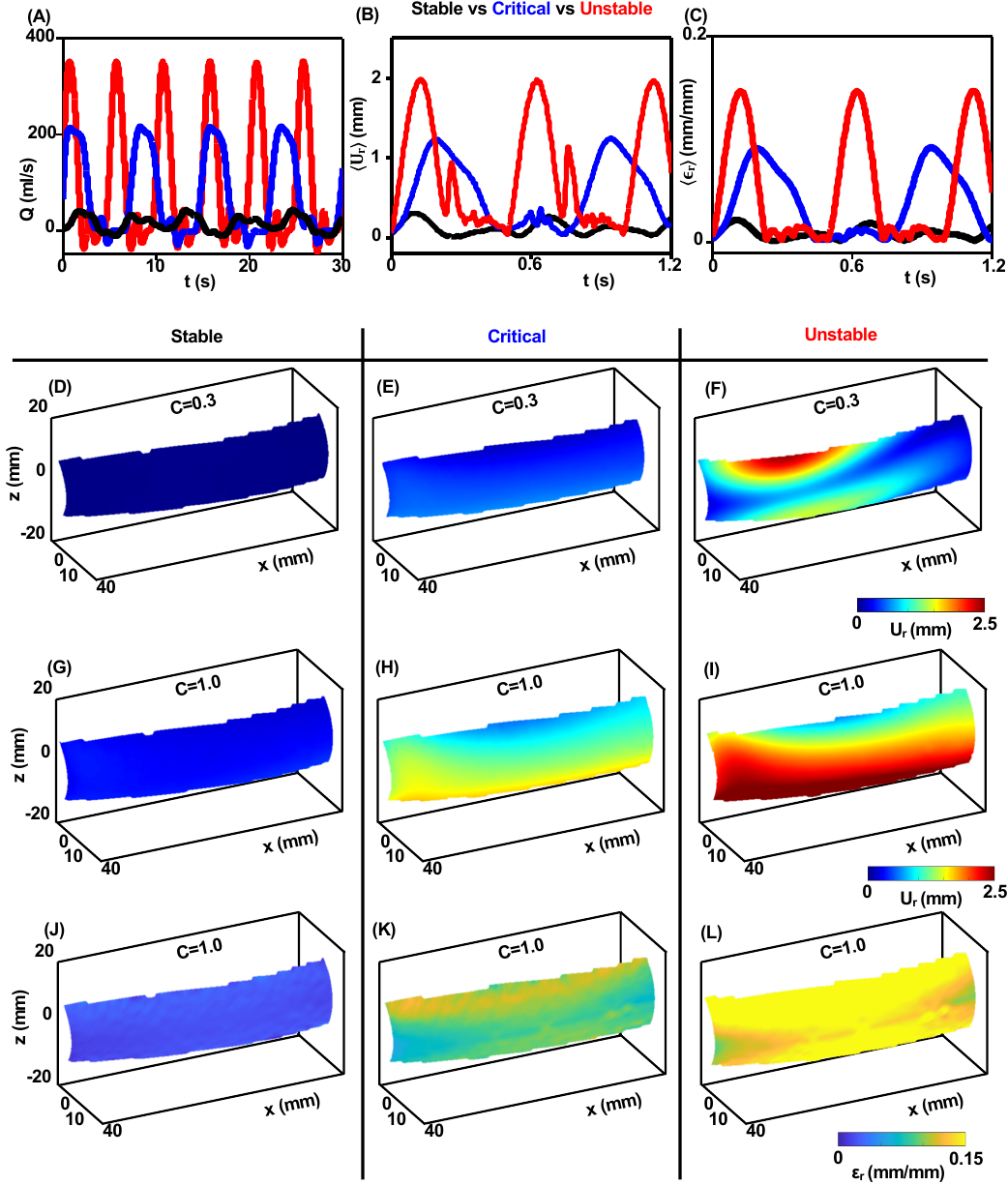}
\caption{The stable $N_{\omega,\text{sp}}=-1.587<0$, critical $N_{\omega,\text{sp}}=0.448\approx0$, and unstable $N_{\omega,\text{sp}}=10.569>0$ cases are visualized here for completeness. The flow rate (A), mean radial deformation (B), and mean radial strain (C) profiles clearly show the higher order temporal modes present for the unstable case vs the smooth profile of the stable scenario. (D,E,F) display the deformation distribution at the same relative point $C$ in each case's respective cycle, such that the dimensional time is $CT$ where $T$ is the period for that case. Note that the unstable scenario exhibits strong localization of deformation patterns. (G, H, I) and (J, K, L) represent the deformation and strain respectively for $C=1$ in all three cases.} 
\label{fig: results2}
\end{figure}

\begin{figure}[]
\centering
\includegraphics[width=10cm]{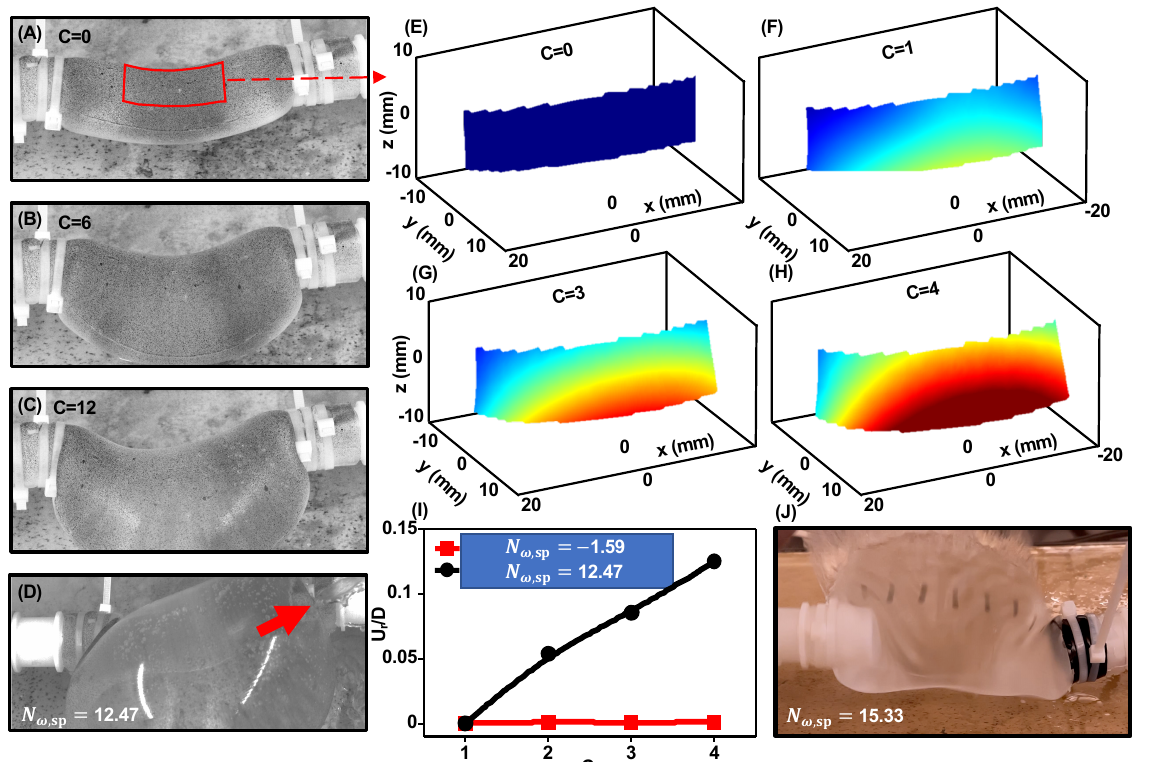}
\caption{The rupture cases are visualized for completeness. (A, B, C, D) show tube expansion at multiples $C$ of the system period $T$ before pinhole rupture occurs (D) at the red arrow. This corresponds to $N^{\text{rupture, 1}}_{\omega,\text{sp}}=12.47$. Deformation maps (E, F, G, H) exhibit the rapid buildup of plastic behavior (I) at the end of each cycle, demonstrating that the leadup to rupture is irreversible. The second rupture scenario associated with $N^{\text{rupture, 2}}_{\omega,\text{sp}}=15.33$ exhibits a characteristic slit failure near the 5 mm length scale observed in vivo \cite{2019Okada, 1955Parkhurst}.} 
\label{fig: rupture_results}
\end{figure}

\begin{figure}[H]
\centering
\includegraphics[width=10cm]{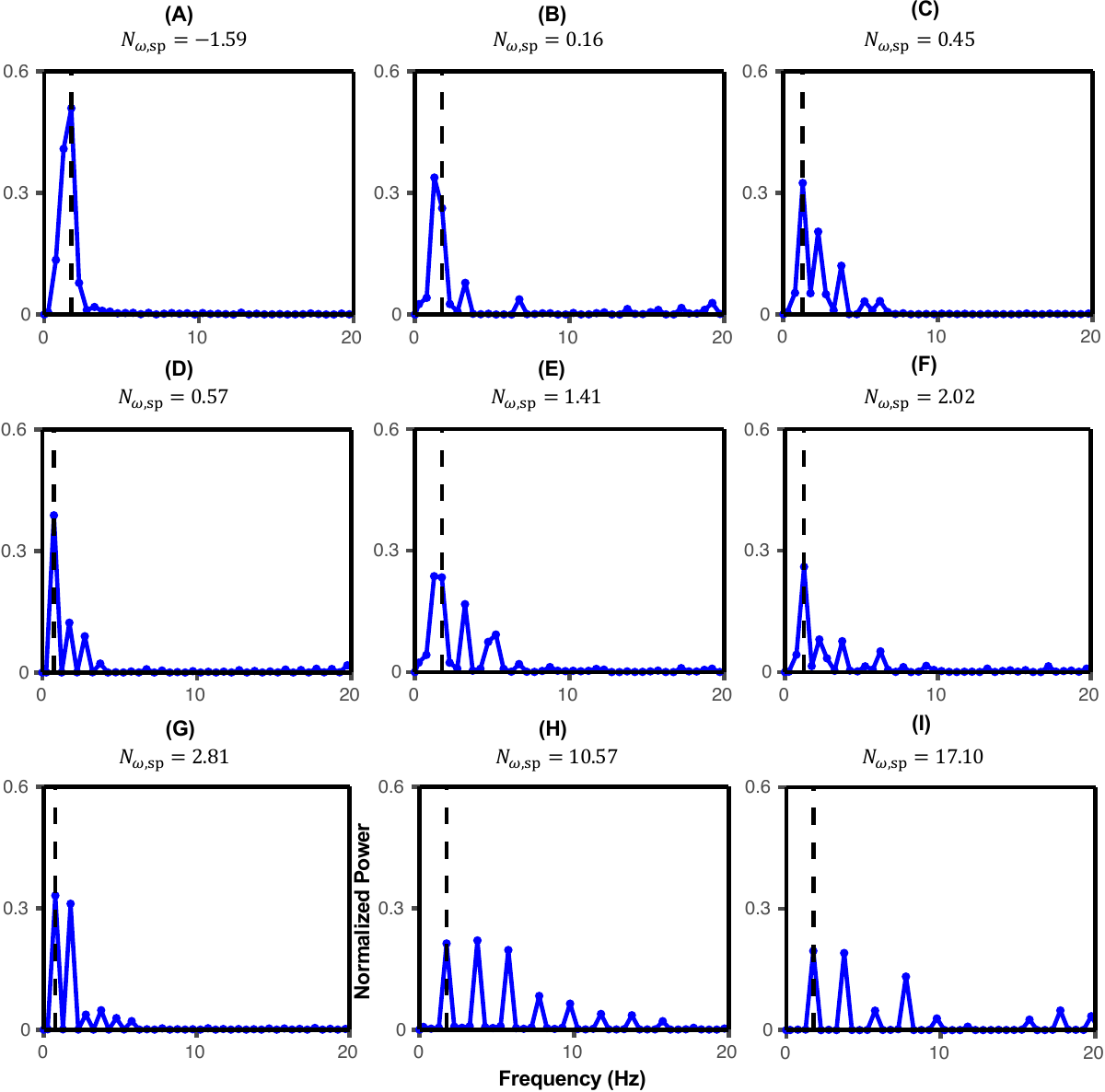}
\caption{Representative graphs of the power vs temporal frequency (Hz) for 9 DIC (Digital Image Correlation) cases. Power is normalized by summed powers of the fundamental frequency and all larger modes. They illustrate the transition from stable flow to fluid-structure instability via the inception and growth of harmonic and subharmonic peaks. This transition occurs as the flutter instability parameter increases from negative to positive values.} 
\label{fig: DIC_panel}
\end{figure}

\begin{figure}[H]
\centering
\includegraphics[width=10cm]{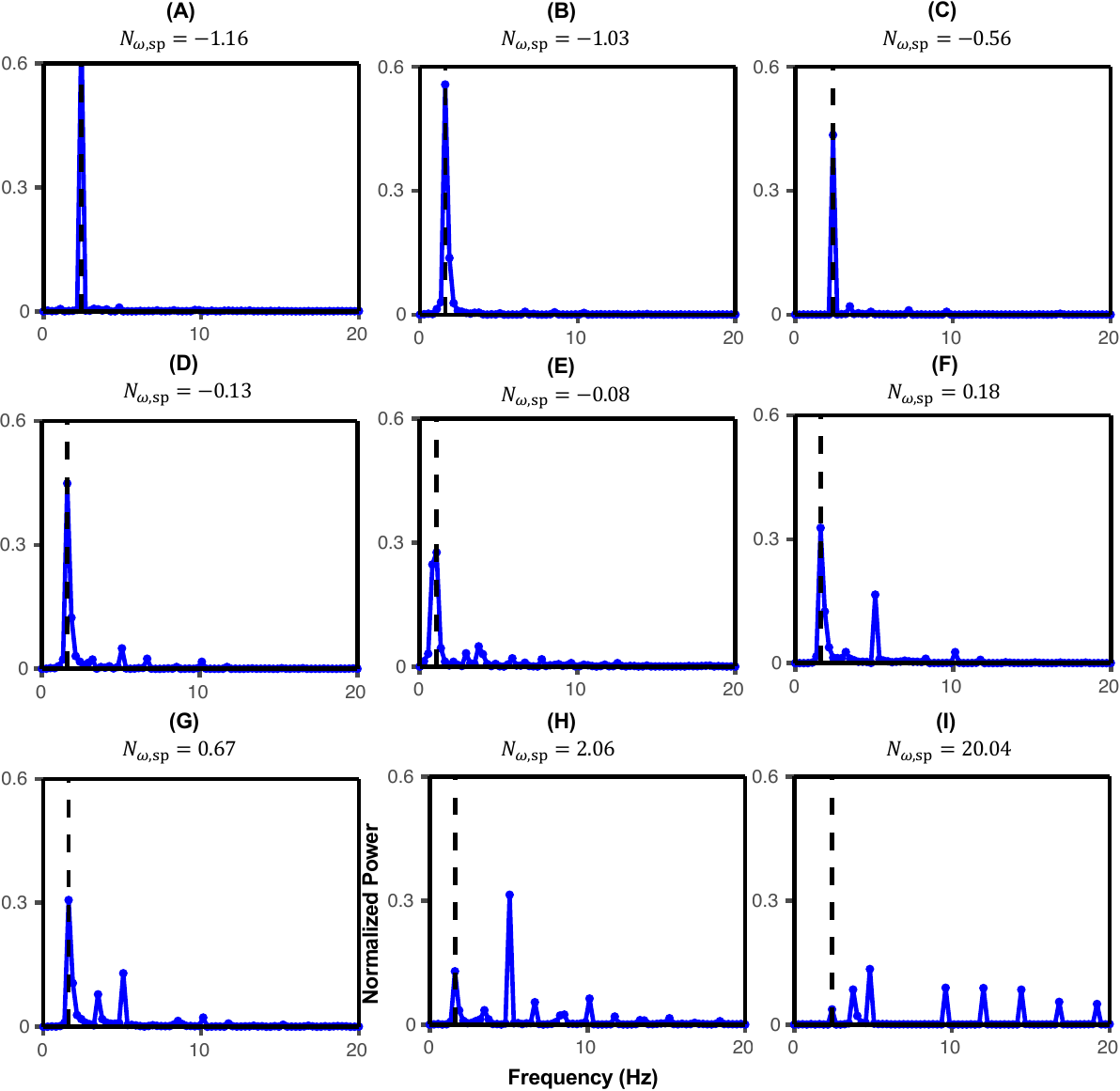}
\caption{Representative graphs of the power vs temporal frequency (Hz) for 9 PTV (Particle Tracking Velocimetry) cases. Power is normalized by summed powers of the fundamental frequency and all larger modes. They illustrate the transition from stable flow to fluid-structure instability via the inception and growth of harmonic and subharmonic peaks. This transition occurs as the flutter instability parameter increases from negative to positive values.} 
\label{fig: PTV_panel}
\end{figure}

Although there are experimental cases with $N_{\omega,\text{sp}}>10$ that have not ruptured (Fig. \ref{fig: permdef}), we note that intense instabilities associated with large aneurysm physiomarker values above $N_{\omega,\text{sp}}\approx 10$ induce strong spatial strain concentration, drive response to high temporal frequency modes, and in summary provide the sufficient condition for rupture to occur. Additionally, the rupture scenarios resulted when the number of cycles were deliberately increased from the three cycles used to measure the temporal and spatial harmonic ratios as well as the aneurysm physiomarker. 

\section{Limitations}

The major limitation of this work is the difference between the behavior and material properties of biological arteries in vivo and that of these synthetic arteries in vitro. 

We have sought to address the difference in materials by selecting PDMS, which like the arterial wall is known to be hyperelastic and also experience viscoelastic creep. Additionally, the stiffness of the PDMS tubes were fabricated to match the pulse wave velocity in the physiological human range (2 to 10 m/s). Note however, that the arterial wall is constructed of multiple layers including the intima, media and adventitia, each of which comprises a heterogeneous population of adipocytes, collagen, elastic, smooth muscle cells, etc. This heterogeneity is not present in the synthetic arteries used in our model, as PDMS is isotropic and homogeneous. 

Nonetheless, the pulse wave velocity measured via magnetic resonance imaging and incorporated in our theory is an emergent mechanical property of the entire aortic wall, effectively encoding the physical response of how these heterogeneous components interact as a whole. The present experiments captures the effective, holistic mechanical behavior of the arterial wall in response to pulsatile forcing at various pressures. We do not expect these in vitro results to exhibit high fidelity in regards to the long time behavior of the aorta, especially given the prominent role of biochemical pathways that induce wall remodeling. However, the short term strain and deformation fields that arise from fluid-structure interactions are dominated by the mechanical forcing and mechanical response experienced by the system. 

In this context, the transition from mechanical stability to instability accompanies a sharp shift in power to higher temporal frequency modes and precedes the significant spatial strain concentration to 1 mm length scales. This jump in mechanical behavior of the synethic arteries observed in vitro is likely to echo a similar jump in aneurysm dilatation for human arteries in vivo. In fact, we have shown this to be true in prior work \cite{2021Zhao}. 
\subsection{Dissection}
The minimum length scale measured for aortic wall tears arising from aneurysm dissection is 1 mm. It is uncertain whether this lower bound is set by data availability, imaging resolution, or the physics of the problem as we have shown. Nonetheless, clinical data reveals that the minimum size of these intimal tears is not capped at a higher value, say, 5 to 10 mm. 
\subsection{Rupture}
Only two rupture cases were explored in this study to avoid damaging the mechanical pump and cameras. For instance, a third case for the 30:1 ratio synthetic artery actuated at 105 bpm was aborted due to significant back pressure causing the mechanical pump to knock against the enclosure. A systematic study of the rupture phase space by varying additional physiological properties comprising the aneurysm physiomarker (predominantly heart rate driven, pressure driven, etc.) would be of great interest. Such a phase diagram would also demonstrate whether pinhole failure always occurs at lower values of the aneurysm physiomarker compared to slit rupture; these procedures would put undue stress on the current experimental setup and remain outside the scope of the present study.

Lastly, note that the thresholds for aneurysm dissection and rupture measured in-vitro may not be easily or even ethically reproducible in-vivo. For instance, aneurysm rupture occurs spontaneously, quickly, and irreversibly. These events often occur outside the hospital\cite{2012Gawenda}, which contributes to the high mortality rate. Since nearly $25\%$ of patients who reach a hospital die before treatment, and $40\%$ of patients who undergo surgery die, there may be no realistic interval during which imaging can be performed to calculate the aneurysm physiomarker. 

\section{Conclusion}
This work uses experiment to prove the existence of a pulsatile fluid-structure instability predicted previously by theory \cite{2021Zhao}. It also demonstrates that instability onset is accurately signaled by the flutter instability parameter $N_{\omega,\text{sp}} = 0$ derived from ab-initio linear stability analysis of the conservation laws. 

On the clinical side, we have shown that permanent arterial dilatation can result from low levels of instability and may depend on the history of instability experienced. Transitioning to instability, then restoring the system back toward stable flow can still induce nonzero permanent deformation that can contribute to the steady dilatation of small aneurysms, even in the absence of material remodeling. 

We have demonstrated that aneurysm dissection appears to be associated with an intermediate instability regime, which for the synthetic arteries begins at flutter instability parameter $N_{\omega,\text{sp}}\gtrsim 2$. This regime introduces strong strain localization at a length commensurate with the smallest scale  intimal tears observed in dissecting aortas. 

Lastly, we have observed that aneurysm rupture can occur in a severe instability regime, which for the the synthetic arteries begins at $N_{\omega,\text{sp}}\gtrsim 10$. Final failure of the aneurysm initiated via two modes that have been also observed clinically in ruptured aortic aneurysms. 

These findings motivate the clinical investigation of the aneurysm physiomarker in diagnosing and treating patients with or at risk of aneurysm formation. The thresholds for the aneurysm physiomarker leading to growth, dissection and rupture can be used to both forecast these different stages as well as intercept them by managing the underlying aortic flutter driving aneurysm progression.

\section{Competing interests}
The authors report no relationships that could be construed as a conflict of interest.

\section{Acknowledgements}
Research reported in this publication was supported by the National Heart, Lung, And Blood Institute of the National Institutes of Health under Award Number F32HL162417. The content is solely the responsibility of the authors and does not necessarily represent the official views of the National Institutes of Health.



\beginsupplement
\section{Supplementary Information}

\end{document}